\documentclass[aps,prb,twocolumn,epsfig,twoside,a4paper]{revtex4}
\usepackage{graphicx} 
\begin{document}

\title{Mode coupling behavior of a Lennard Jones binary mixture:
a comparison between bulk and confined phases}

\author{A.~Attili$^\dagger$, P.~Gallo$^\dagger$\footnote[1]
{Author to whom correspondence  should be addressed; e-mail: 
gallop@fis.uniroma3.it} and M.~Rovere$^\dagger$,}
\address{$\dagger$ Dipartimento di Fisica, 
Universit\`a ``Roma Tre'', \\ 
and Democritos National Simulation Center,\\ 
Via della Vasca Navale 84, 00146 Roma, Italy.}

\begin{abstract}
We present a quantitative comparison at equivalent thermodynamical
conditions of bulk and confined dynamical
properties of a Lennard Jones binary mixture upon
supercooling. Both systems had been previously found to
display a behavior in agreement with
the Mode Coupling theory of the evolution of glassy dynamics.
Differences and analogies of behavior are discussed focusing in particular on 
the role of hopping in reducing spatially correlated
dynamics in the confined system with respect to the bulk. 

\end{abstract}


\date{\today}
\maketitle


\section{Introduction}

The structural and dynamical properties of bulk 
supercooled liquids are now relatively well understood 
in the range close to the crossover temperature of the
Mode Coupling Theory of the evolution of glassy dynamics, 
MCT.~\cite{goetze}
This theory describes, in its ideal version, 
the relaxation mechanism of supercooled liquids 
as mastered by the cage effect. The motion of atoms
or molecules is restricted by the presence of nearest neighbors
that surround and trap the tagged particle forming a cage around it.
When the cage relaxes, due to cooperative motion, the particle
diffuses. Relaxation times of the cage diverge on approaching
a temperature $T_C$ where,
in this framework, the liquid undergoes a structural arrest. 
In most liquids this dynamical singularity is avoided
since close to $T_C$ the hopping relaxation channel opens. 
The temperature of glass transition is reached when 
also hopping is frozen. In this context 
$T_C$ remains an important temperature that marks 
the ideal crossover from a regime mastered by the cage effect to 
one mastered by hopping effects. This crossover temperature can
be estimated through an MCT analysis of dynamical properties of the liquid
both with experiments and computer simulations.

Relaxation modes described by MCT are cooperative in character.
Cooperativity is a basic concept in several 
theories and models of the glass transition.
The existence of cooperative rearranging regions, 
CRR, intimately related to dynamic 
heterogeneities, 
can be a possible interpretative mechanism 
for the approach to the glass transition.~\cite{donati}
The extent of the CRR is expected to increases with
decreasing temperature.  
In the presence of a restricted environment the behavior
of these regions might be modified and this might imply 
a change of the dynamical behavior.  
Confinement can therefore offer possible clues 
for a better understanding of general mechanisms for the glass 
transitions in the bulk. 

To assess the validity
and the possible limits of applicability of MCT in confinement
can help to bridge the concepts of hopping, caging and 
dynamical heterogeneities. These phenomena appear in fact 
all fundamental ingredients for the glass transition to occur.
Only very few computer simulation
studies on the applicability of MCT on supercooled 
liquids in confined geometries are 
available.~\cite{krakoviack,baschnagel,waterconf,kobconf} 
A detailed comparison of the dynamics of a glass former in the bulk and 
in a confined state can give new insights in understanding the 
behavior of the supercooled liquids. 
This comparison can nonetheless be quantitatively carried out only if
it is performed at equivalent thermodynamical conditions.

In this paper we consider one of the most
studied model for glass former liquids, a Lennard Jones binary mixture, LJBM, 
which is known to test MCT in its bulk phase~\cite{kob}.
We carry on a direct comparison between 
a set of simulations that we have previously published of 
the LJBM confined in a disordered array
of soft spheres~\cite{euro-noi,pella1,pella2} and 
an equivalent bulk phase.
The LJBM hosted by the matrix is in
a strong confining environment where a large fraction of 
particles are interfacial~\cite{euro-noi}. 
This microporous medium 
can be considered a model for real compounds like silica 
xerogels.\cite{rosinberg,monson,page}
The confined liquid has been found in our previous works
to test MCT.~\cite{euro-noi,pella1,pella2}
Recently the mean square displacement results for this system
have also been found to fit in the framework of a 
recently proposed mean field 
theory of the glass transition.~\cite{michio}

A difficulty to face in the comparison with the bulk is  
the presence of variations of the local 
density of the confined LJBM as a function of temperature due to the
soft sphere repulsive potential. There is not a unique way to define 
a free volume.~\cite{pella1}
Therefore, in order to obtain a quantitative  
comparison of the dynamical properties 
we performed here computer simulations of the bulk liquid
along a thermodynamical path where the $(T,P)$ values
follow the isochore obtained for the confined system
upon supercooling. This implies that for the bulk phase
analyzed in this paper 
the approach to $T_C$ is not obtained along isochoric, isobaric
or isothermal paths, as usually done for bulk glass formers.

The paper is structured as follows:
in the next section we present the details of the newly performed bulk
simulations, and a brief description of the previously studied 
confined LJBM.
The third section describes the behavior of the density
correlators in the $\alpha$ relaxation region. The fourth section
explores the role of hopping in bulk compared to confinement.
The fifth section presents an analysis of dynamical
clusters in the two systems. Last section is
devoted to conclusions.

\section{Simulation details}

We report data analysis obtained from
molecular dynamics simulations trajectories of a 
Lennard-Jones 80:20 binary mixture (LJBM)
defined as in ref.~\onlinecite{kob} in the bulk phase.
The LJBM state points are simulated at the same temperatures and pressures
of previously reported data from the same mixture in a confined 
phase~\cite{euro-noi,pella1,pella2}.

In the following the 80\% atoms will be referred to as $A$ and the
20\% as $B$. 

In the confined system the LJBM was  
embedded in an off lattice matrix of soft spheres, labelled with $S$. 
The simulation box of the confined system contained 
a rigid disordered array of $N_S=16$ soft spheres. 
A constant volume with a selected simulation box length $L=12.6$
was used. For further details referring to the confined system
see refs.~\onlinecite{euro-noi,pella1,pella2}.

Temperature and pressure of the bulk system
during cooling and equilibration procedures were 
controlled via a Berendsen weak coupling to 
thermal and pressure baths.
In this case a varying box length ranging from $L=10.34$ to $L=9.36$ 
upon decreasing temperature has been used.

For each thermodynamical point, molecular dynamics simulation to evaluate
static and dynamical properties,
have been then performed in the microcanonical 
ensemble for both bulk and confined systems. 

The confined system was studied for temperatures 
ranging from $T=5$ to $T=0.37$~\cite{euro-noi,pella1,pella2}
while the bulk range of temperature investigated for comparison
goes from $T=5$ to $T=0.48$. $T=0.37$ and $T=0.48$ represent the lowest 
temperatures at which we are respectively 
able to equilibrate the confined and bulk 
system. 

We used a timestep of 0.01 for $T > 0.53$ and $0.02$ for
$T< 0.53$ for the confined system
and $0.01$ for $T > 0.8$ and $0.02$ for
$T< 0.8$ for the bulk system. 
The total production time 
of the lowest temperature investigated, $T=0.37$,  was of 
$t_{\rm run}=14$ millions of timesteps. These values would correspond
for liquid Argon to $T=44.3$ $K$ and $t_{\rm run}=4.2$ $\mu s$.

\begin{figure}[h!]
\includegraphics[width=8.5cm]{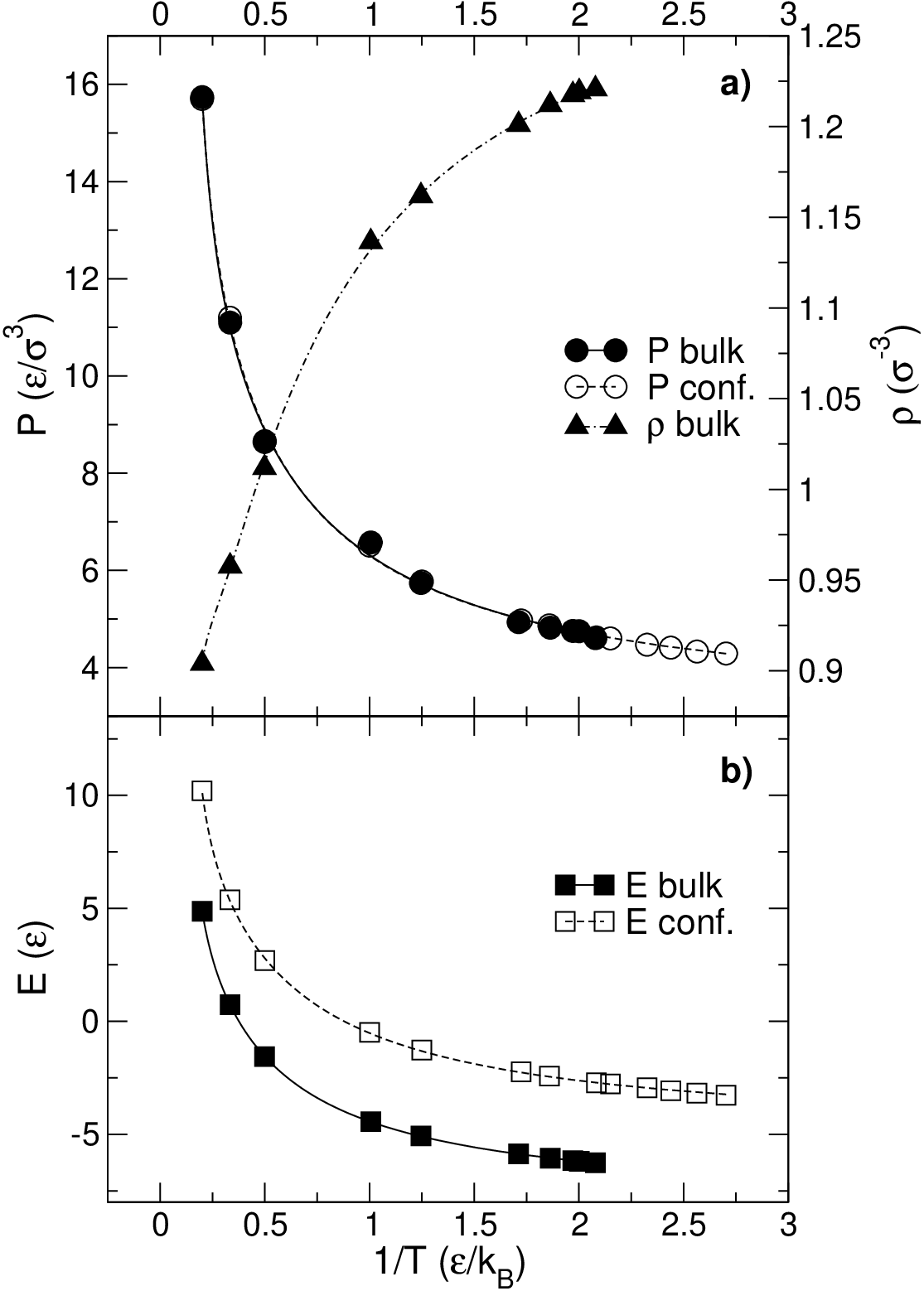}
\caption{a) Pressure as a function of inverse temperature 
of both bulk and confined~\protect\cite{pella1}
systems; the corresponding bulk density 
is also shown. 
b) Total energy per atom versus inverse temperature for 
bulk and confined~\cite{pella1} systems. In both panels, lines depict 
simple polynomial interpolations.}
\protect\label{fig:1}
\end{figure}

The behavior of total energy, pressure and, for the bulk, density as
a function of temperature is depicted in Fig.~\ref{fig:1}.
Thermodynamic $(T,P)$ points of the bulk follow the isochoric path found
upon supercooling for the confined system~\cite{pella1}. 
In the same panel we note also that upon supercooling, 
in the range of investigated temperatures, an increase
of number density $\rho$ occurs in the bulk system where $\rho=0.9 \div 1.2$. 
In the confined system the density cannot be exactly calculated
due to the presence of the
repulsive soft sphere potential. However a qualitative 
estimate of the density of the confined mixture carried 
out with Voronoi tessellation  has been reported in ref.~\onlinecite{pella1}. 
The range of densities
estimated with this method for the confined system was found to be 
$\rho=0.7 \div 1.2$, similar to that found in our equivalent bulk liquid. 

In the lower panel of Fig.\ref{fig:1} we also reported the values of the total
equilibrium energies per atom of the two systems.
A similar trend is observed.
As expected the energy of the confined system is shifted
above the bulk one due to the additional
term of interaction with the soft spheres in the potential energy.
At higher temperatures an average increase of about 
$5.0$ $\epsilon_A$
of the internal energy of the confined system is found. 
This difference tends to reduce asymptotically to
$3.0$ $\epsilon_A$ upon cooling.
Over the range of temperatures and densities investigated the
bulk and the confined liquids show no signatures of phase separation
as we can deduce from the smoothness of the curves. 
This is consistent with a study 
of the liquid limits of the bulk LJBM.~\cite{sriprl}.

We found that the main features of the pair correlation functions and 
of the static 
structure factors of our bulk LJBM (not reported) are similar to those of the
confined liquid \cite{pella1}.

\section{The alpha-relaxation region}

The slowing down of the dynamics 
of a supercooled liquid on approaching $T_C$ can 
be characterized by analyzing the temperature dependence of the 
single particle or self intermediate scattering function 
$F^{(s)}(Q,t)$, SISF, for both types of particles.
In this section we report the bulk results on the SISF analysis
compared to the previously obtained results 
for the confined system.~\cite{euro-noi,pella1,pella2}

\begin{figure}[b!]
\includegraphics[width=8.0cm]{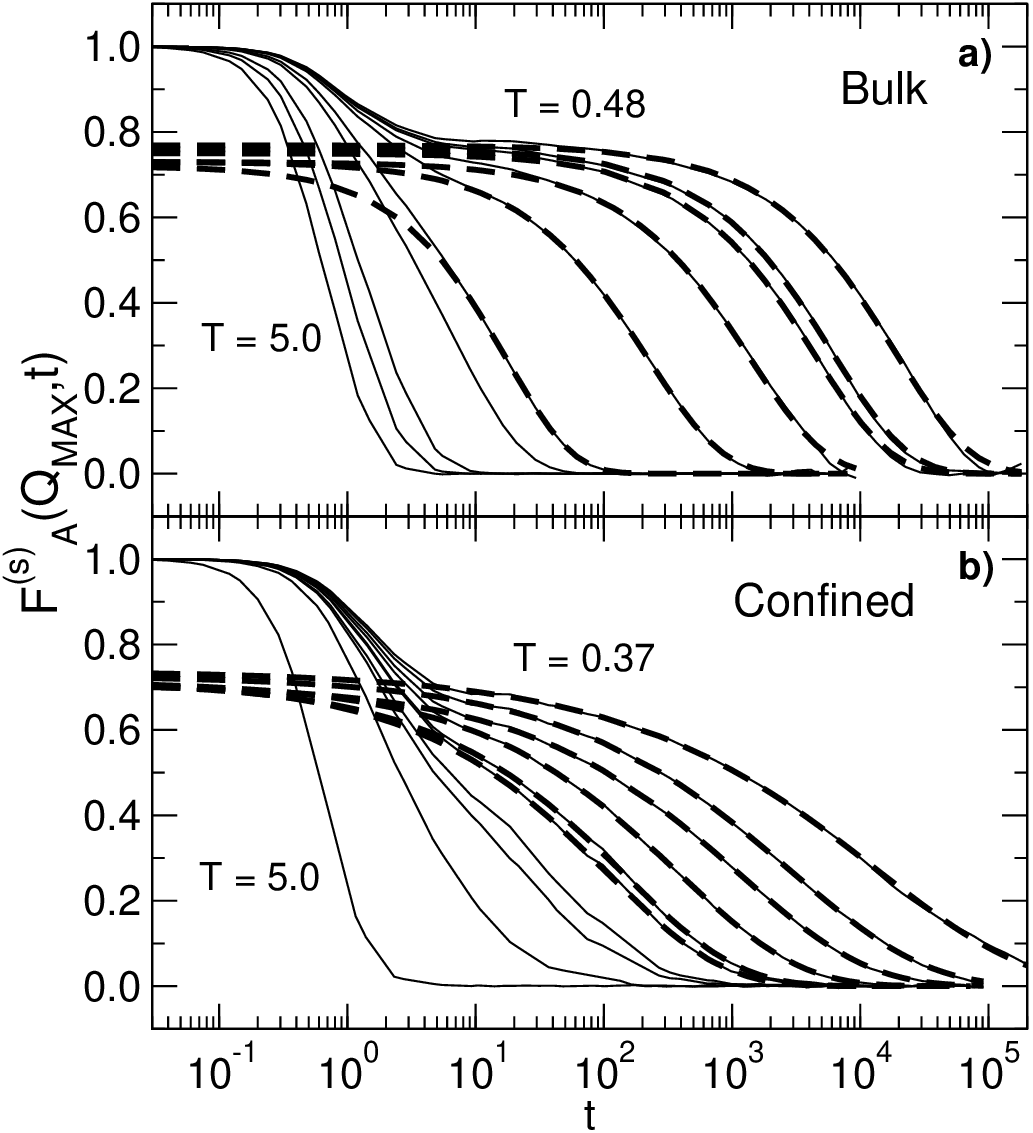}
\caption{Self part of the intermediate scattering function for the $A$ 
particles, $F^{(s)}_A(Q_{\rm MAX},t)$, for different temperatures 
for the bulk (a) 
and the confined (b)~\cite{pella2} systems. The  correlators
are evaluated at $Q_{\rm MAX}=7.06$, the position of the first 
peak in the static structure factor for the $A-A$ correlation. 
The dashed lines are fits to KWW stretched exponential 
(see Eq.~\ref{eq2}), in the region of the late $\alpha-$relaxation.}
\protect\label{fig:3}
\end{figure}

\begin{figure}[tb!]
\includegraphics[width=8.5cm]{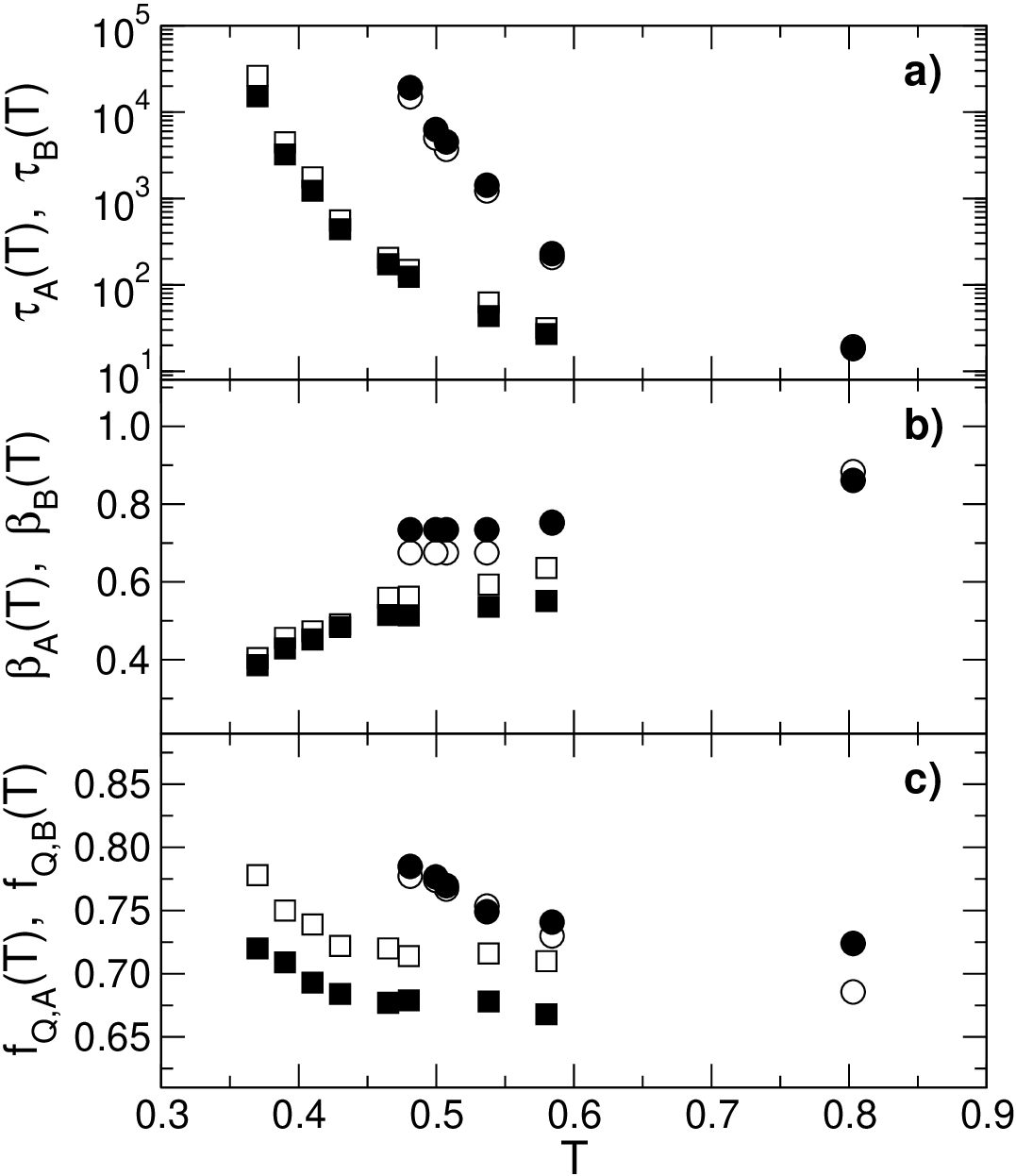}
\caption{Fitting parameters of the KWW curves plotted in Fig.~\ref{fig:3} 
for both bulk and confined systems as functions of the temperature.
Circles refer to bulk 
A (filled) and B (open) particles. Squares
refer to confined~\cite{pella2} A (filled) and B (open) particles. 
a) Relaxation times $\tau_{A}(T)$ and $\tau_{B}(T)$. b) Stretching 
parameter $\beta_A(T)$ and $\beta_B(T)$. c) Form factor 
$f_{QA}(T)$ and $f_{QB}(T)$.}
\protect\label{fig:4}
\end{figure}

The temperature dependence of the bulk SISF is analyzed 
in the following for wave lengths corresponding to the inter-particle spacing, 
i.e. for  $Q=Q_{\rm MAX}(T)$, located at the first peak 
of the static structure factors. 
Around these values of $Q$ MCT features are best 
evident.~\cite{goetze}
Since the positions of the diffraction peak show a weak $T-$dependence
in order to pursue consistently our comparison
we use for the bulk analysis 
the same values of $Q_{{\rm MAX},A}=7.06$ and 
$Q_{{\rm MAX},B}=5.90$ that we used in the confined 
LJBM~\cite{pella2}.

The results for $A$ particles are reported in Fig.~\ref{fig:3}
for both bulk and confined~\cite{pella2} systems on the same scale for 
a close comparison. Similar results are found 
for $B$ particles (not reported). From the figure we recognize 
upon supercooling for both bulk and confined mixtures
the onset of the MCT predicted two-step relaxation scenario.
At $T_C$ in the idealized version of MCT this scenario leads
to the bifurcation of the long time limit of the 
correlator. This limit also called non ergodicity parameter,
is zero for the liquid and jumps to a finite value on crossing the 
MCT critical line.
The existence of the plateau is related to the MCT two time fractals.
The dynamical regime that covers the approach 
to the plateau of the correlator is referred in literature as 
the $\beta-$relaxation regime. 
This regime is  followed by the $\alpha-$relaxation 
that starts when the correlator departs from the plateau.
For long times, in the late $\alpha-$relaxation regime, 
the intermediate scattering 
function decays to zero in a non-exponential fashion. 

\begin{figure}[b!]
\includegraphics[width=8.5cm]{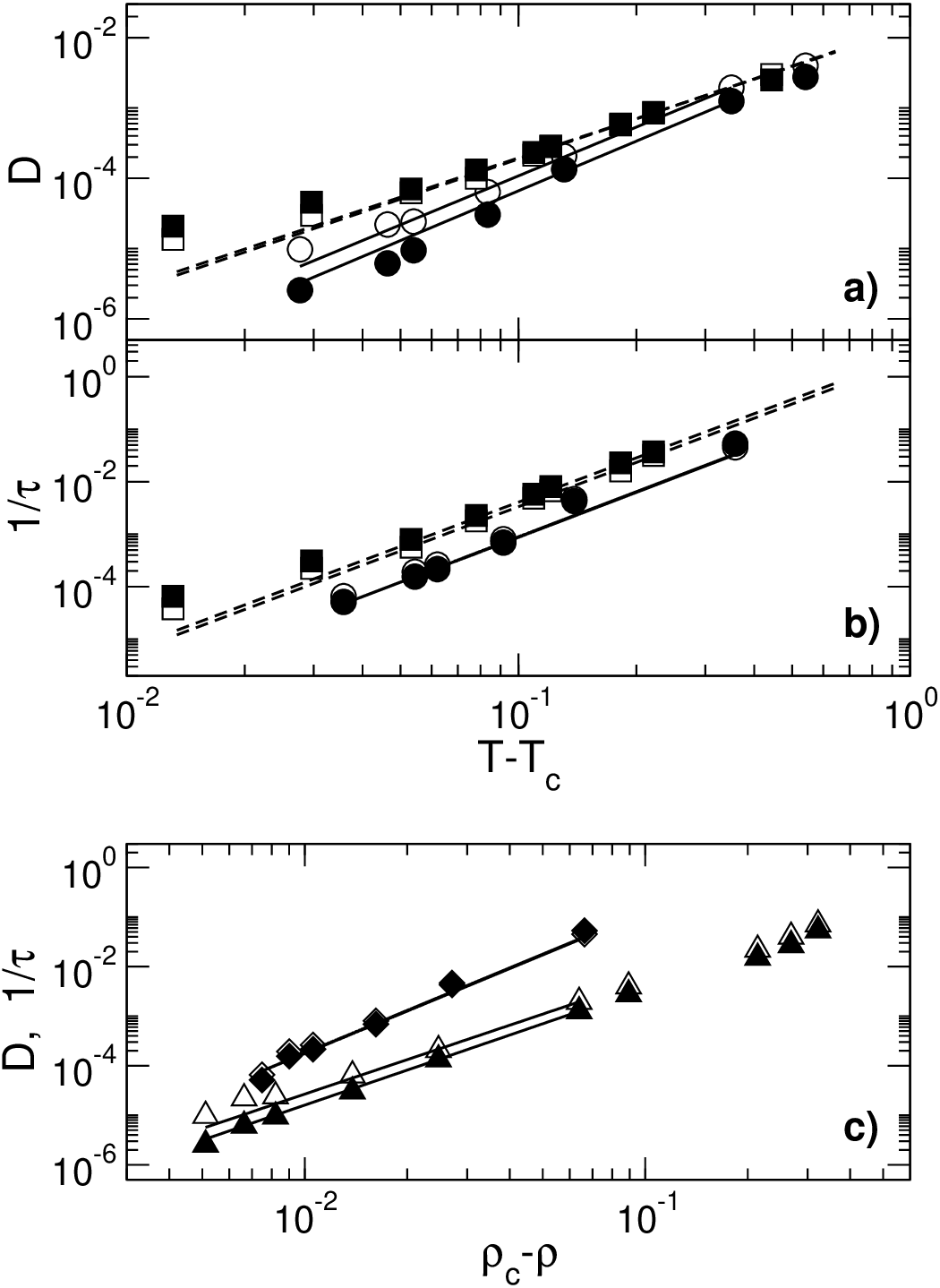}
\caption{Power law fits of the diffusion coefficients (a) and of the inverse 
relaxation times (b) as a function of temperature for confined (squares) 
~\protect\cite{pella2} and bulk (circles) liquids. 
In (c), only for the bulk case,
analogue power law behavior of $D$ (triangles) and $\tau$ (diamonds)
as functions of density is shown.
In all panels filled symbols refer to $A$ particles, open symbols to 
$B$ particles. 
The extrapolated MCT parameter, $T_C$ and $\rho_C$, and exponent $\gamma$, 
are reported in Tab.~\ref{tab:1}.}
\protect\label{fig:5}
\end{figure}

\begin{table}[ht!]
\begin{center}
\begin{tabular}{||c||c|c||c|c||}
\hline
\hline
\multicolumn{1}{||c||}{} & \multicolumn{2}{c||}{Bulk Mixture} 
& \multicolumn{2}{c||}{Conf. Mixture} \\
\cline{2-5}
 & $A$ & $B$ & $A$ & $B$\\
\cline{1-5}
\cline{1-5}
\multicolumn{5}{||l||}{(from $D$)} \\
$T_C$ & $0.453$ & $0.453$ & $0.356$ & $0.356$ \\
$\rho_C$ & $1.226$ & $1.226$ & $-$ & $-$ \\
$\gamma$ & $2.348$ & $2.309$ & $1.860$ & $1.890$ \\
\multicolumn{5}{||l||}{(from $\tau_{\alpha}$)} \\
$T_C$ & $0.445$ & $0.445$ & $0.356$ & $0.356$ \\
$\rho_C$ & $1.228$ & $1.228$ & $-$ & $-$ \\
$\gamma$ & $2.849$ & $2.987$ & $2.80$ & $2.80$ \\
\hline
\hline
\end{tabular}
\caption{Power-laws fit 
parameters of bulk and confined~\protect\cite{pella2} mixtures, 
for $A$ and $B$ atoms. In the upper half of the table are reported 
values obtained from the analysis of diffusion constants $D$, while 
in the lower part values obtained from the analysis of relaxation times 
$\tau_{\alpha}$.}
\label{tab:1}
\end{center}
\end{table}

We note that the shape
of the correlators in bulk and confined liquids differs.
In particular at the lowest temperatures investigated, 
$T=0.48$ and $T=0.37$ respectively for bulk and confined systems,
the extension of the plateau rapidly grows
spanning up to over three decades in the bulk, while it is much less
defined and covers only
one decade in the confined system. After the plateau the correlator
decays much faster to zero in the bulk than in the confined system
where long time tails related to a more consistent
stretching of the relaxation law can be observed.

The $T-$dependence of the $\alpha-$relaxation time $\tau_\alpha(T)$, 
can be extracted from the fits of the $F^{(s)}(Q,t)$ to the 
Kohlrausch-Williams-Watts (KWW) stretched exponential function
\begin{equation}
\phi_\alpha(t)=f_{Q}(T) 
\exp\left[ -(t/\tau_\alpha(T))^{\beta(T)} \right],
\label{eq2}
\end{equation}
where $\beta(T)$ is the stretching parameter. 
$f_{Q}(T)$ can be considered
as an effective Lamb M\"ossbauer factor and is 
related to the height of the plateau.~\cite{goetze} 
In the ideal version
of the theory this height coincides with the 
non ergodicity parameter  for $T\le T_C$, 
$f_{Q}(T)=\lim_{t\to\infty}\phi_\alpha(t)$.~\cite{goetze}

Although the KWW functional form is not an exact solution of MCT equations, 
it has been shown that it is able to reproduce remarkably well the behavior 
of the numerical solution evaluated from these equation, in the 
$\alpha-$relaxation regime.~\cite{goetze}

The results of the fit are reported,
together with the density correlators, in Fig.~\ref{fig:3} 
where we can see that the agreement 
of the fitting function to the correlators curves is very 
good. 

Fitting parameters for both particle types in bulk 
and confined~\cite{pella2} systems as functions of the temperature, 
are shown in Fig.~\ref{fig:4}.

As expected for systems
testing MCT, the relaxation times $\tau_{A}(T)$ and $\tau_{B}(T)$
increase dramatically as the temperature is 
lowered, varying up to four order of magnitude in the $T-$ranges investigated.
Similar trends in bulk and in confinement~\cite{pella2} are 
observed but we note that the relaxation times for the confined system are 
consistently smaller than 
those of the bulk. This is a clear indication of a faster dynamics 
due to repulsive confinement. 

The departure from the Debye exponential relaxation is quantified 
by the stretching parameters $\beta_A$ and $\beta_B$. 
These can be connected to the presence of heterogeneity in the dynamics, 
due to the coexistence of dynamical regimes characterized 
by a distribution of relaxation times. 
The stretching parameters of both systems have values close to $1$ 
(simple Debye relaxation) at high temperatures. 
MCT asymptotically predicts these parameters to have a weak 
temperature dependence. 
This prediction is satisfied in the bulk liquid, 
where, upon decreasing the temperature, the parameter is seen to lower 
and to reach a constant plateau value close to 0.8, whereas, in the case 
of the confined system, the parameter does not reach a constant value, 
but instead keeps on decreasing regularly
down to 0.4 at the lowest temperature investigated~\cite{pella2}. 
We will see in the next section that the latter behavior can be related 
to the presence of 
significant {\it hopping} processes found in the confined system 
at these temperatures.

The $T-$dependence of the parameter $f_Q$ 
(height of the plateau) is rather weak as predicted by MCT
and has a similar trend in both confined~\cite{pella2} and bulk systems. 
We found $0.70\le f_Q \le 0.77$ for the bulk with similar values 
for different species. For the confined system different 
values for $A$ and $B$ particles were found: $0.66\le f_Q \le 0.72$ for $A$ 
particles and $0.72\le f_Q \le 0.77$ for $B$ particles. 

\begin{figure}[tb!]
\includegraphics[width=8.5cm]{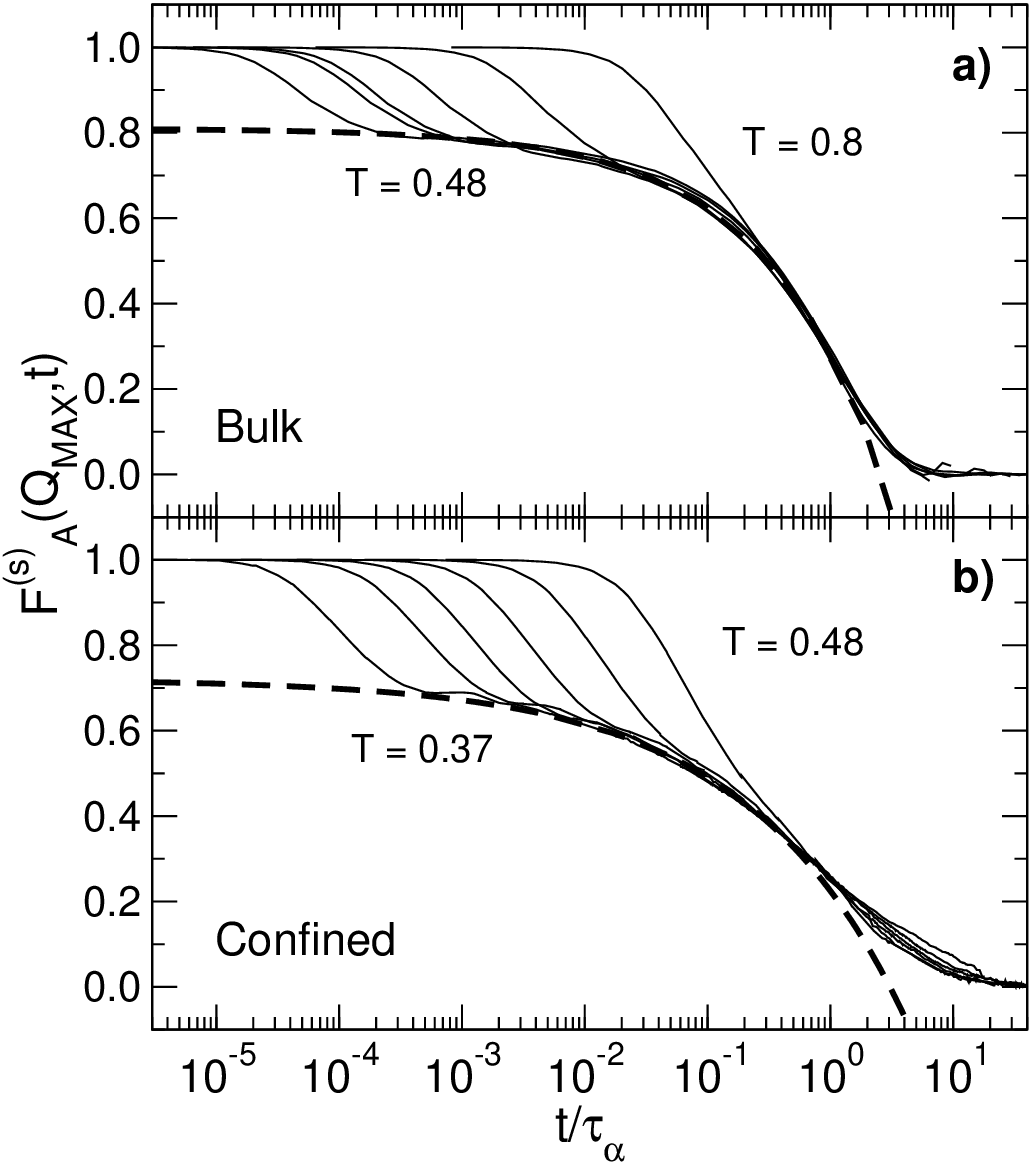}
\caption{Self part of the intermediate scattering function for the $A$ 
particles, $F^{(s)}_A(Q_{\rm MAX},t)$, with $Q_{\rm MAX}=7.06$, for different 
temperatures, scaled by $\tau_\alpha$ (obtained from a KWW fit, see 
Fig.~\ref{fig:4}); the dashed curve is the fit to a von Schweidler law 
defined in Eq.~\ref{eq:4}, (see Tab.~\ref{tab:2}). In (a) are presented 
bulk correlators at temperatures $0.48 \le T \le 0.8$, in (b) confined 
correlators at temperatures $0.37\le T \le0.43$.~\protect\cite{pella2}}
\protect\label{fig:6}
\end{figure}

\begin{table}[ht!]
\begin{center}
\begin{tabular}{||c||c|c||c|c||}
\hline
\hline
\multicolumn{1}{||c||}{} & \multicolumn{2}{c||}{Bulk Mixture} & 
\multicolumn{2}{c||}{Conf. Mixture} \\
\cline{2-5}
 & $A$ & $B$ & $A$ & $B$\\
\cline{1-5}
\cline{1-5}
$f^c_Q$ & $0.81$ & $0.81$ & $0.72$ & $0.80$ \\
$h_Q$ & $0.52$ & $0.53$ & $0.35$ & $0.43$ \\
$b$ & $0.450$ & $0.531$ & $0.355$ & $0.340$ \\
\hline
$r^b_s$&$0.197$&$0.216$&$0.246$&$0.222$ \\
$r^{MSD}_s$&$0.179$&$0.220$&$0.234$&$0.232$ \\
\hline
\hline
\end{tabular}
\caption{Von Schweidler fit 
parameters of bulk and confined~\cite{pella2} mixtures, 
for $A$ and $B$ atoms.}
\label{tab:2}
\end{center}
\end{table}

According to the ideal version of MCT, the relaxation times, 
$\tau_\alpha$, as well as the diffusion coefficients, $D$, 
are expected to follow asymptotically 
a power-law divergence as functions of temperature and density
near the cross-over point $(T_C,\rho_C)$
\begin{equation}
D,\tau_{\alpha}^{-1} \propto (T-T_C)^{\gamma}, 
\quad D,\tau_{\alpha}^{-1} \propto (\rho_C-\rho)^{\gamma},
\label{eq:5b}
\end{equation}
where $T_C$ and $\rho_C$ are respectively the cross-over temperature and 
density of the theory, and $\gamma$ is the exponent parameter. 
The fits 
to power-laws are reported in Fig.~\ref{fig:5} overlayed to $\tau_{\alpha}$ 
and $D$ data found in both confined~\cite{pella2} and bulk systems. 
The coefficients $D$ are extracted from the 
mean square displacements (MSD), not reported, in the diffusive regime.  

We found good agreement of the relations of Eq.\ref{eq:5b} 
with the data in the range
 $0.48\le T \le 0.80$ ($1.16\le \rho \le 1.22$) for the bulk to be compared to
$0.41\le T \le 0.58$ obtained for the 
confined system~\cite{pella2}. These values
correspond, in terms of the small parameter of the theory 
$\epsilon=(T_C-T) /T_C$, to a relative reduction of 
the ideal MCT range of applicability in 
confinement of about $30\%$ with respect to the bulk.

Results for the fitting parameters are reported in table \ref{tab:1}.
$T_C$, $\gamma$
and (for the bulk only) $\rho_C$, are similar for the different 
species $A$ and $B$ as predicted by the theory. 
The same crossover temperature is found from $D$ and $\tau$.
In the confined liquid $T_C\simeq 0.36$ is sensibly lower 
than in the bulk  $T_C\simeq 0.45$. 

The critical density of the bulk is slightly higher than the 
density simulated for the lowest temperature bulk,
and also very close to the estimate of the density of the lowest 
temperature in the confined system, see Sec. II.

We note that the values of $\gamma$ extracted from $\tau_\alpha$
are similar for both bulk and confined mixtures, $\gamma\sim 2.8$.
Contrary to the MCT predictions, the values of $\gamma$
extracted from $D$ differ from those extracted from $\tau_\alpha$ both in bulk
and in confined systems. 
This difference is more marked in the presence of confinement and a possible 
source of these discrepancies is the presence of hopping
that will be discussed in sec. IV. 

The {\it time-temperature superposition principle}, TTSP, predicted by MCT 
to hold asymptotically near $T_C$, states that the shape 
of the correlator curves in the late $\beta-$relaxation and 
early $\alpha-$relaxation time regimes does not depend on temperature. 
Consequently the intermediate scattering functions can be written as
\begin{equation}
F^{(s)}(Q,t)=\hat{\phi}_Q(t/\tau_\alpha(T))
\label{eq:TTSP}
\end{equation}
where $\tau_\alpha$
is a time scale associated to the $\alpha-$relaxation 
decay of the correlation function and $\hat{\phi}_Q$ is a master function. 
We used in the scaling of the SISF  
the $\tau_\alpha(T)$ obtained via KWW fits.

The correlators are plotted for different temperatures versus 
rescaled time $t/\tau_\alpha(T)$ in Fig.~\ref{fig:6} for $A$ particles. 
The intermediate scattering 
functions reported in the figure are evaluated in the same temperature 
ranges used in the analysis of power-law divergence, i.e.
$\epsilon\sim 0.4$ for the confined mixture~\cite{pella2} and 
$\epsilon\sim 0.7$ for the bulk one. From the figures it is seen 
that also for the bulk with varying density
the rescaled correlator curves fall on top of each other within the 
the late $\beta-$relaxation region up to the $\alpha-$relaxation time range. 
Similar results hold also for $B$ particles (not shown).

MCT theory predicts also a precise functional form of the master function 
$\hat{\phi}(t)$ in the $\beta-$relaxation regime. The form is a 
von Schweidler (VS) power law,
\begin{equation}
\hat{\phi}(t)=f^c_Q-h_Q(t/\tau_{\alpha})^b.
\label{eq:4}
\end{equation}
Best fits to our master plots 
are included in Fig~\ref{fig:6} (dashed lines). 
Rescaled time intervals used in the fits are $10^{-3}\le t/\tau_{\alpha} 
\le 10^0$.
The values of the free parameters, $f^c_{Q}$, $h_{Q}$ and $b$ are reported 
in Tab.~\ref{tab:2}. As it can be seen from the figures, there is a good 
agreement with the data while
we observe that the prediction of MCT that b should be the same
for both A and B particles is verified only in confinement.
The better agreement with MCT of the 
confined system might be related to packing constraint induced by the 
confined geometry and the repulsive potential that 
further reduce the tendency of the mixture to phase separation.
As already observed for the stretching parameter the von
Schweidler exponent b lowers in confinement reflecting a larger 
distribution of the relaxation times.
The critical non ergodicity parameter $f^c_Q$ offers an estimate of
the mean square displacement $r_s^2$ of the particles at a time
$t=1/\omega_c$, where $\omega_c$ is the cutoff frequency below which
the whole alpha peak is located, i.e. the time where the $\alpha$ relaxation
starts. In fact:
\begin{equation}
\ln f^c_Q= \frac{-(r_sQ)^2}{3}+O(Q^4)
\label{eq:NEP}
\end{equation}
The values of the $r_s$ extracted both from 
Eq.~\ref{eq:NEP} are consistent with the ones obtained 
from the plateau of the 
MSD (not shown) as reported in Tab.\ref{tab:2}.
We observe that A and B particles in confinement and
B particle in bulk move on similar distances while the packing of
large A particles is stronger in the bulk.

We note that the VS law has been tested to be valid for an extended range
of $Q$ around the peak of the structure factor for both confined
and bulk systems \cite{kob,pella2}, not shown here.

\begin{figure}[h!]
\begin{center}
\includegraphics[clip,width=8.5cm]{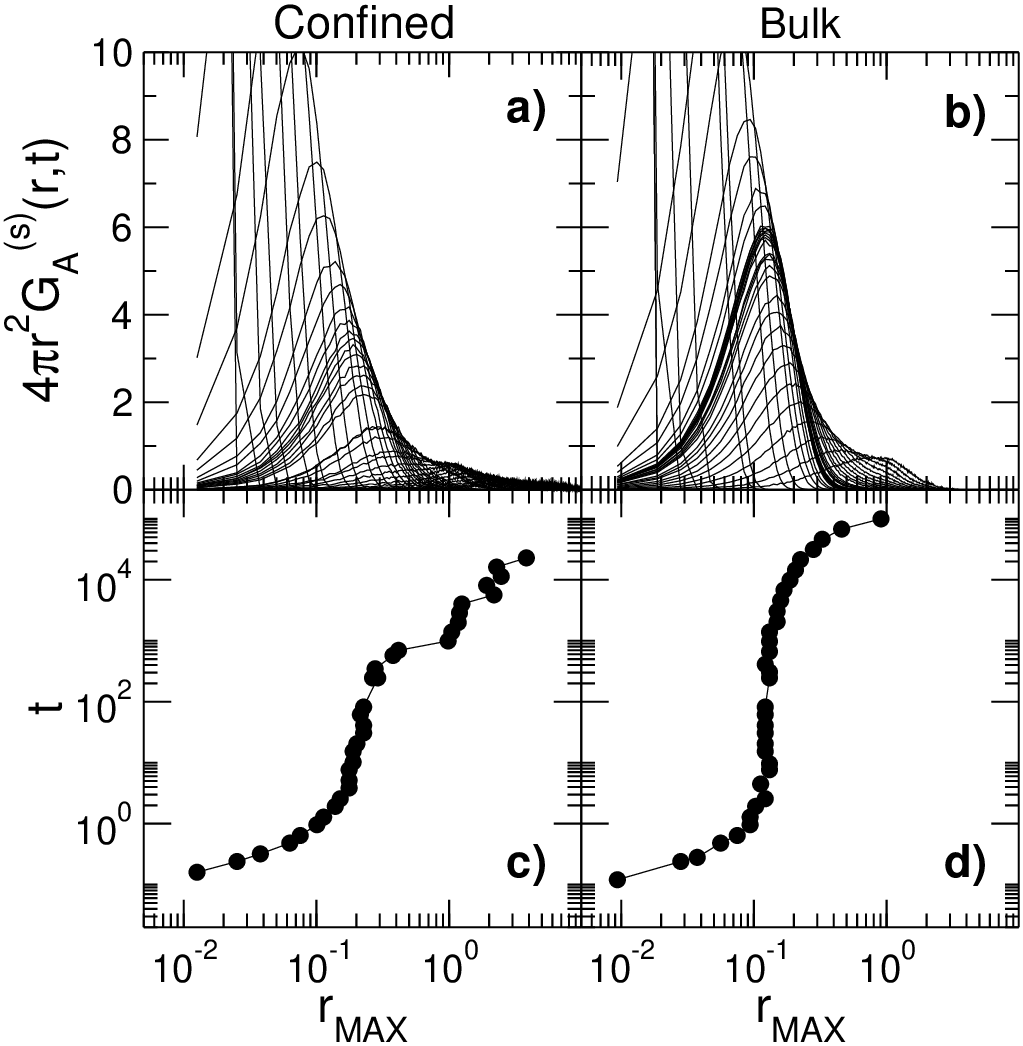}
\end{center}
\caption{Self part of the van Hove function (for $A$ particles)
evaluated at $T=0.48$ for the confined (a) and the bulk
(b) systems. The range of time used is $2 \times 10^{-2} \le t \le 10^5$ for
the bulk and $2 \times 10^{-2} \le t \le 2\times10^4$ for the confined
systems.
In panels (c) and (d) are displayed, in a double logarithmic scale,
the positions of the peaks in the van Hove function (x axis), for
confined and bulk systems respectively, versus times (y axis).}
\protect\label{fig:9}
\end{figure}

\begin{figure}[t]
\begin{center}
\includegraphics[clip,width=8.5cm]{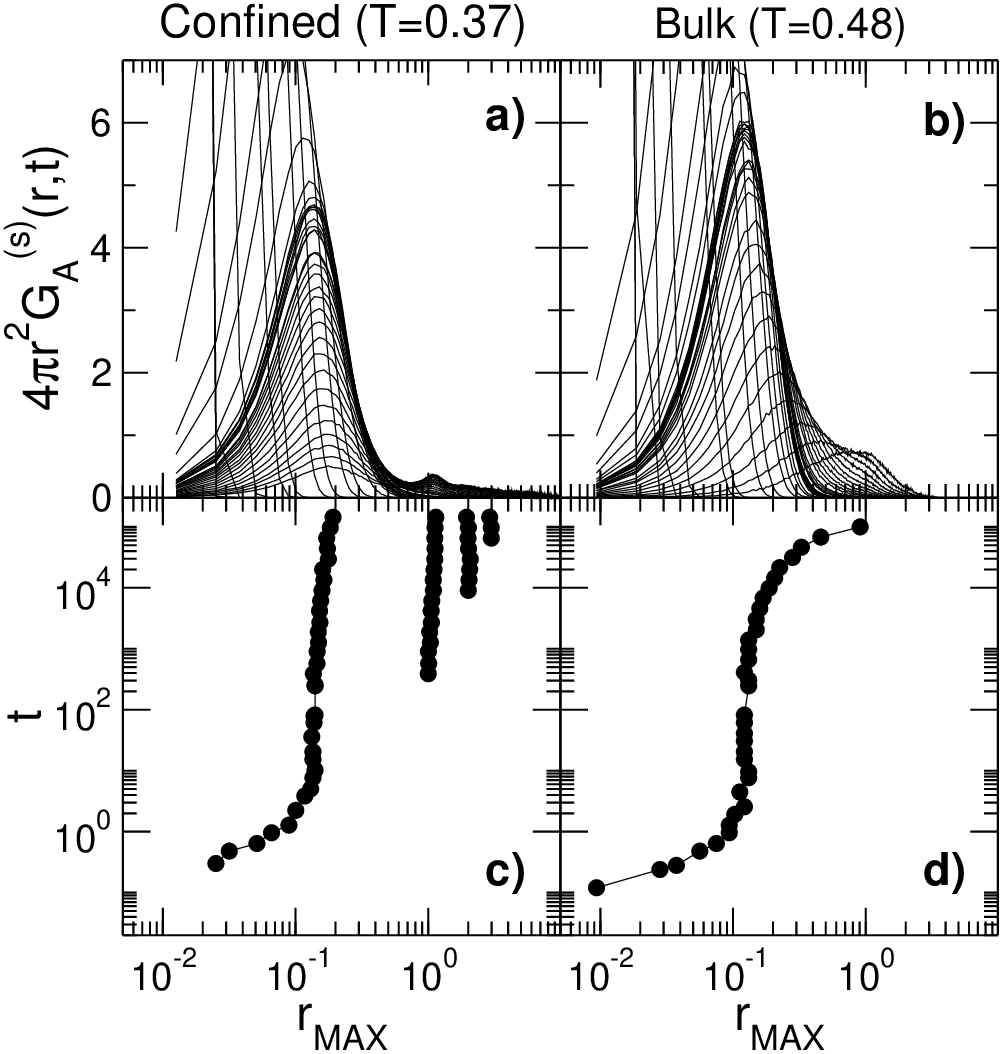}
\end{center}
\caption{Self part of the van Hove function (for $A$ particles)
evaluated at $T=0.37$ for the confined (a) and $T=0.48$ for the bulk
(b) systems. The range of time used is $2 \times 10^{-2} \le t \le 10^5$ for
the bulk and $2 \times 10^{-2} \le t \le 2\times10^4$ for the confined
system.
In panels (c) and (d) are displayed, in a double logarithmic scale,
the positions of the peaks in the van Hove function (x axis), for
confined and bulk system respectively, versus times (y axis).}
\protect\label{fig:9b}
\end{figure}

\begin{figure}[h!]
\begin{center}
\includegraphics[width=9cm]{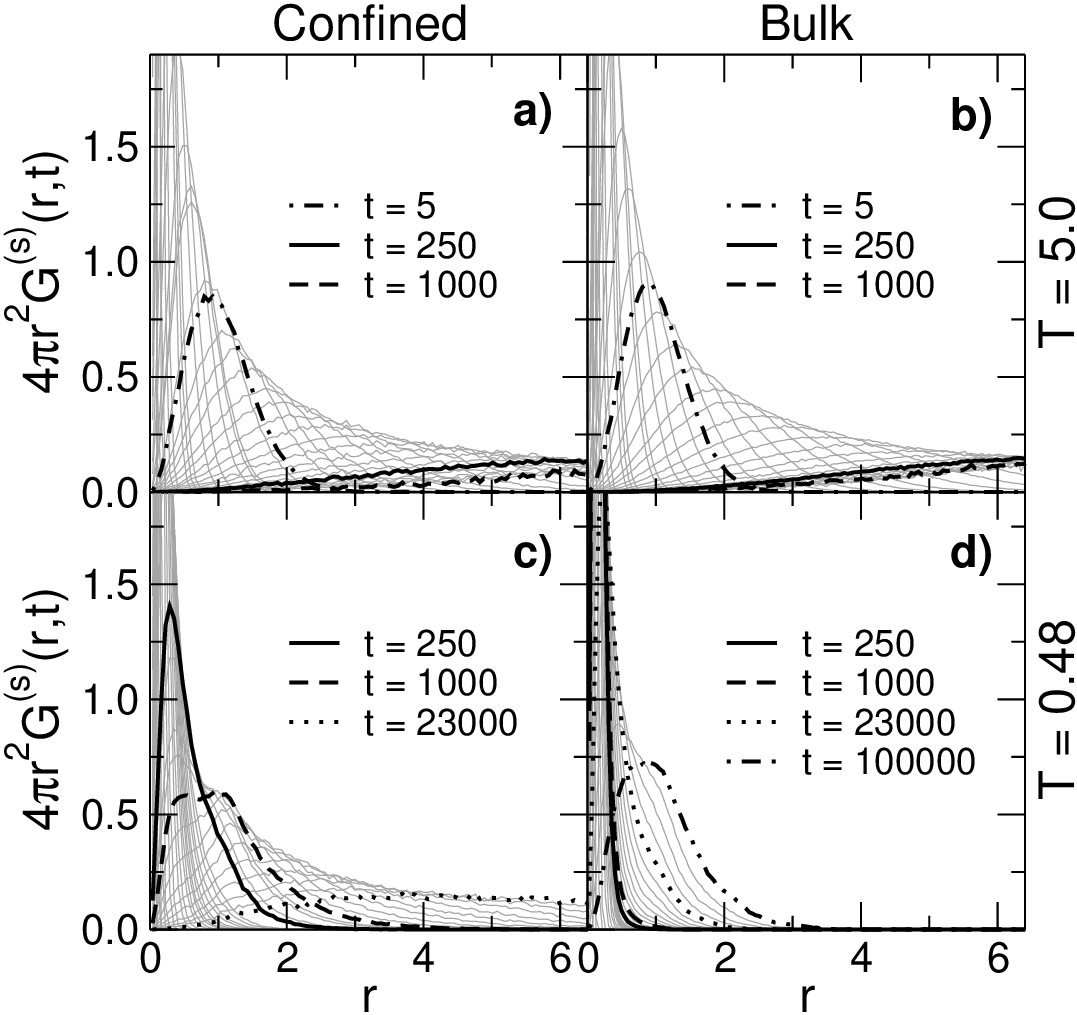}
\end{center}
\caption{Comparison between bulk and confined systems for the self 
part of the van Hove correlation function $G^s(r,t)$ for A particles, 
evaluated at a high ($T=5.0$) and at a low temperature ($T=0.48$). 
Curves belonging to different times are displayed in the same plot, 
some of these curves are plotted with thicker lines. The four panels, 
adopting the same scales, represent $G^s(r,t)$ for: a) confined system, 
$T=5.0$; b) bulk system, $T=5.0$; c) confined system, $T=0.48$; d) bulk 
system, $T=0.48$.}
\protect\label{fig:10}
\end{figure}

\begin{figure}[h!]
\includegraphics[width=9cm]{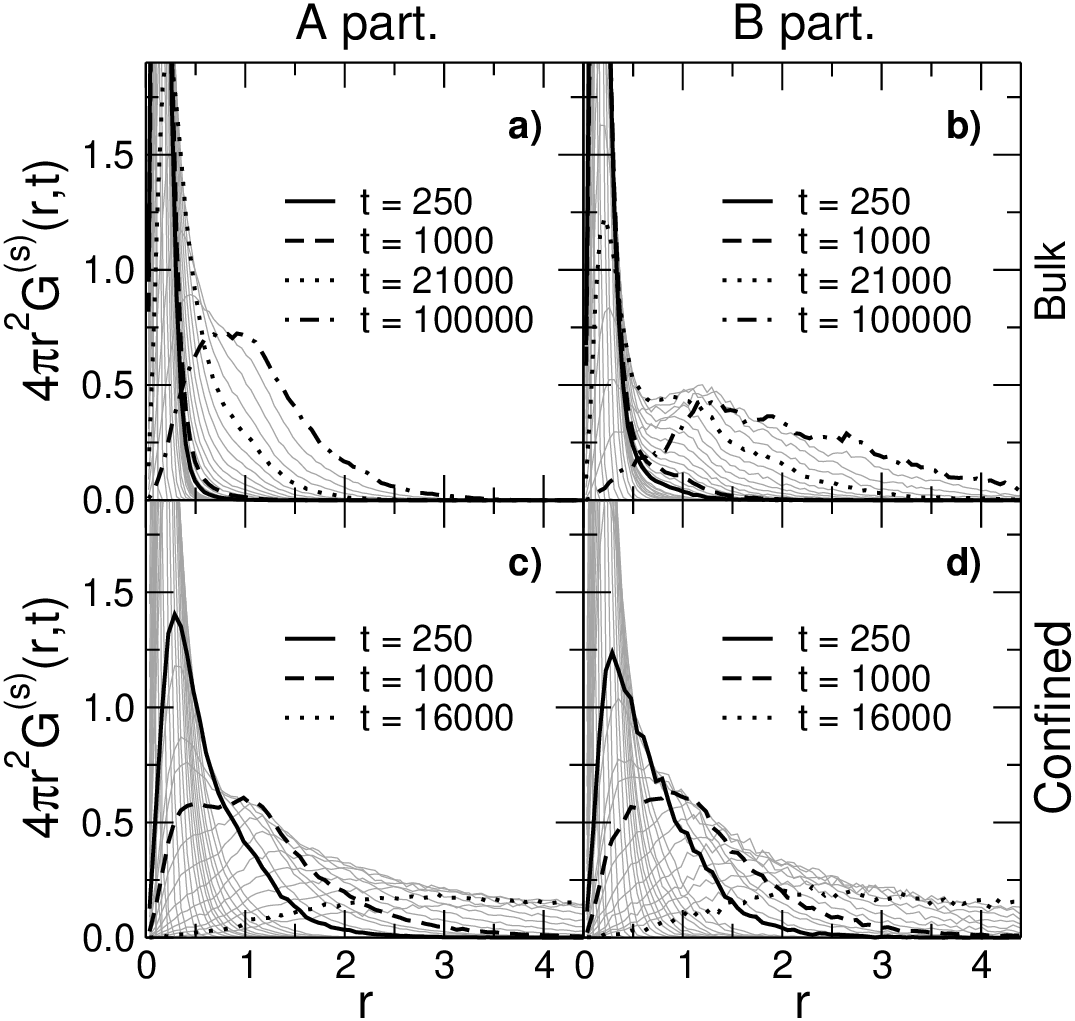}
\caption{Comparison between bulk and confined systems and between 
different species of particles, for the self part of the van Hove 
correlation function $G^s(r,t)$. The correlators are evaluated at the 
same temperature, $T=0.48$, and for different times, some of these evidenced 
with thicker lines. The four panels, adopting the same scales, represent 
$G^s(r,t)$ for: a) bulk system, A particles; b) bulk system, B particles; 
c) confined system, A particles; d) confined system, B particles.}
\protect\label{fig:11}
\end{figure}

\begin{figure}[h!]
\begin{center}
\includegraphics[clip,width=8.5cm]{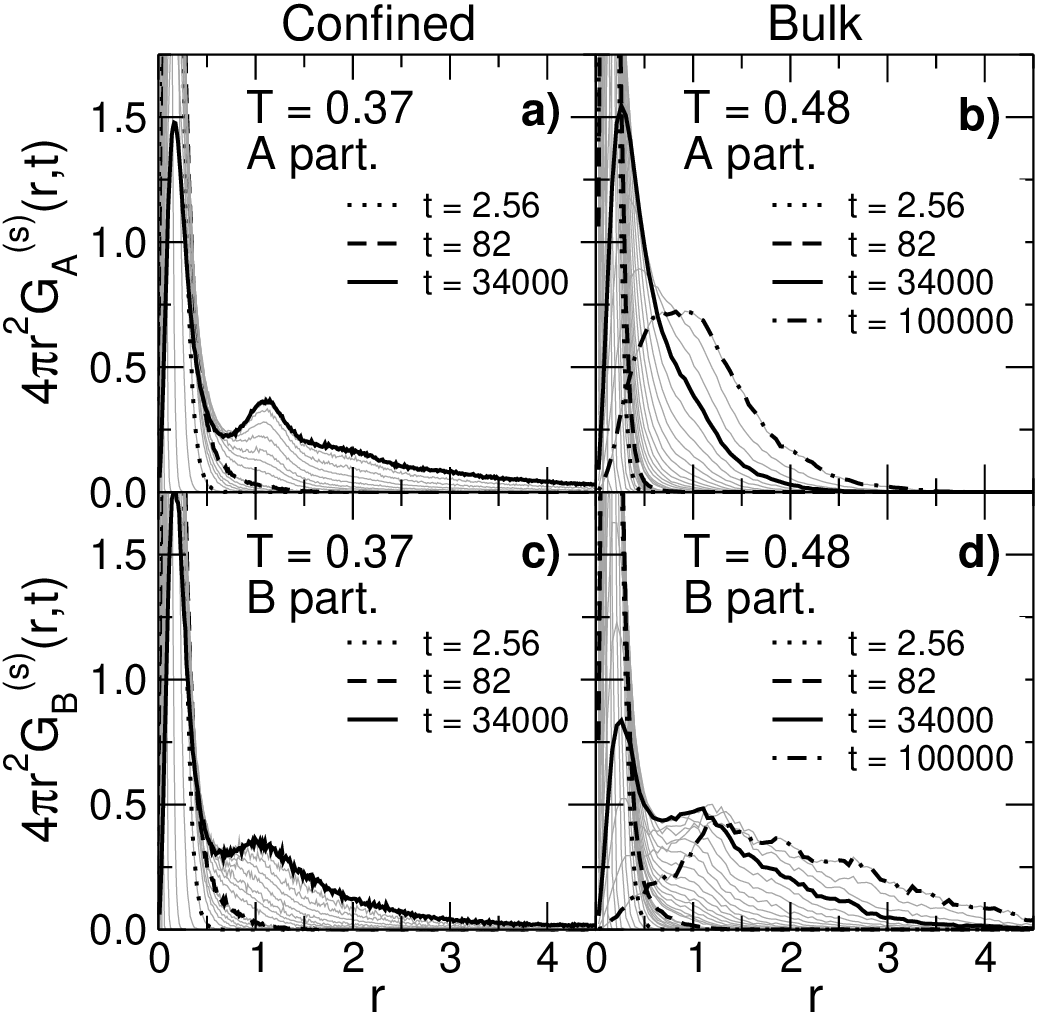}
\caption{Comparison between bulk and confined~\cite{pella1} systems 
and between 
different species of particles, for the self part of the van Hove 
correlation function $G^{(s)}(r,t)$. The correlators are evaluated 
at the same value of the small parameter
$\epsilon=0.04$ of MCT, which correspond to temperatures $T=0.48$ 
for the bulk liquid and $T=0.37$ for the confined system. 
The correlator is shown in the time space region where hopping 
is visible, i.e. the late $\beta-$relaxation regime. 
The functions are normalized to unity. The four panels, 
adopting the same scales, represent $G^s(r,t)$ for: 
a) confined system, $A$ particles; b) bulk system, 
$A$ particles; c) confined system, $B$ particles; 
d) bulk system, $B$ particles.}
\protect\label{fig:12}
\end{center}
\end{figure}

\section{Hopping effects}

\subsection{van Hove Correlation Function}

To investigate the relevance of hopping effects above $T_C$ in our systems 
we now investigate the 
distribution of the average distances. This can be done by means of the 
self part of the van Hove correlation function (VHCF) defined as
\begin{equation}
G^{(s)}_\mu(r,t)=\frac{1}{N_\mu}\sum_{i_\mu=1}^{N_\mu}\left< 
\delta({\bf r}_{i_\mu}(t)-{\bf r}_{i_\mu}(0)-{\bf r}) \right>,
\end{equation}
This function is related to the intermediate scattering 
function via Fourier transform and 
$4\pi r^2 G^{(s)}_\mu (r,t)$ is the probability 
density for a particle of specie $\mu$ of being 
displaced of a distance $r$ in a time 
interval $t$.

The van Hove correlator as a function of the distance $r$, has been evaluated 
for different temperatures and for several different times, that span 
from the ballistic to the diffusive regime of the MSD, 
$10^{-2}\le t \le 10^5$, 
with a $t\sim 2^n$ geometric progression. The evaluations have been carried 
out for both $A$ and $B$ particles and for each system.

In Fig.~\ref{fig:9} we report a 
comparison between the distance distributions found in the confined and bulk 
LJBM (panel (a) and (b) respectively). 
In these plots both systems are observed 
at the same temperature $T=0.48$, i.e. 
the lowest temperature simulated for the bulk 
liquid. From these figures we can see that in both systems the distribution 
shows a simple Gaussian shape. The Gaussian peak maxima shift at greater 
distances, when the curves are sampled at increasing delay times.

In both bulk and confined systems the shift follows a quadratic
$t-$dependence  for early times, $0.01 < t < 1$, typical of the
ballistic motion of the  particles. For greater times than those of
the ballistic regime, a time range  where the curves cluster on each
other as predicted by MCT starts to develop. 
This clustering is evidenced in the plots
reported in panel (c) and (d) of Fig.~\ref{fig:9},  where the
positions of the maxima in the VHCF at their corresponding times are
reported in a double logarithmic scale. In the case of the bulk
liquid, the  time range correspond to $2 < t < 3 \times 10^3$, and
the curves shows a tendency  to cluster at the localization radius
$r \sim 0.15$. At the same  temperature, for the confined system we
found a similar but less pronounced  clustering  in the more limited
range $3 < t <2 \times 10^2$ and at $r \sim 0.2$.  
The position of the peak of the
VHCF remains constant for the presence of oscillating particles which
are trapped in their transient cages. The time ranges in which we
observe this trapped dynamics correspond to those of the plateau seen
in the SISF.
We note how at this temperature the effect of trapped dynamics 
is more accentuated in the bulk than in the confined system.

In the study of the supercooled phase in the asymptotic region near $T_c$ 
of MCT, a more significant comparison for the details of the dynamical 
behavior is achieved by considering not absolute temperatures, but 
instead temperatures relative to the critical temperature, i.e. 
thermodynamic conditions with the same value of the small parameter 
of the theory $\epsilon$. This comparison is 
supported by the validity of the TTSP for the supercooled regime. 
In the previous section we demonstrated that this principle is indeed 
satisfied by our systems.

We now compare in Fig.~\ref{fig:9b}, in the same 
fashion, the VHCF for the two lowest 
temperatures simulated in bulk and confinement. These temperatures correspond
to the same distance from the respective $T_C$.
Their value of the small parameter is $\epsilon=0.04$.
An interesting feature is observed for the confined case.
The particles remain trapped in the frozen cage for a time longer than
the bulk. Escape is guaranteed only by hopping. In this correlator
in fact now three hopping peaks are clearly visible for the A particles while 
hopping is totally absent for this kind of particles in the bulk for
the time scales of the figure.

In order to evidence the behavior of the $G^{(s)}_\mu (r,t)$ at higher times
(and higher distances) we show in Figs.~\ref{fig:10}, \ref{fig:11}
and \ref{fig:12}  a four-panel comparison of 
the correlators reported in a linear scale,  between correlation 
functions for $A$ and $B$ atoms, bulk and confined systems.
In  Fig.~\ref{fig:10} the comparison is reported only for A
particles in bulk and confinement at a high temperature and a
low temperature. 
While at high temperature the systems exhibit very similar 
correlators at the same times, larger differences are found at the lower 
temperature, where
the behavior of the van Hove correlators indicates a much greater 
mobility and a more evident presence of hopping
of particles in the confined system than in the bulk.
In Fig.~\ref{fig:11} the comparison is carried out 
at the same low temperature, $T=0.48$, for both kind of particles. 
From the figure it
can be argued that the small $B$ particles have a larger 
tendency to exhibit hopping in the bulk. It is also clear
that confinement succeeds in reducing 
hopping for $B$ particles and to enhance hopping for
$A$ particles. 
The comparison carried out at the same  $\epsilon$
is reported in Fig. \ref{fig:12}.
From this comparison it is even more clear than in the former figures
what is the fundamental difference 
between the dynamics in the bulk and the dynamics in confinement.
In the bulk we have hopping only for B particles. For these particles 
the phenomenon is marked, but in a quite
distinct time range with respect to the time range of the caging.
In confinement for both A and B particles we have a marked hopping 
already when a part of the particles is still rattling in the cage. 
The overall result is that dynamics appears similar
for $A$ and $B$ particles only in confinement.

\begin{figure}[h!]
\begin{center}
\includegraphics[width=8.5cm]{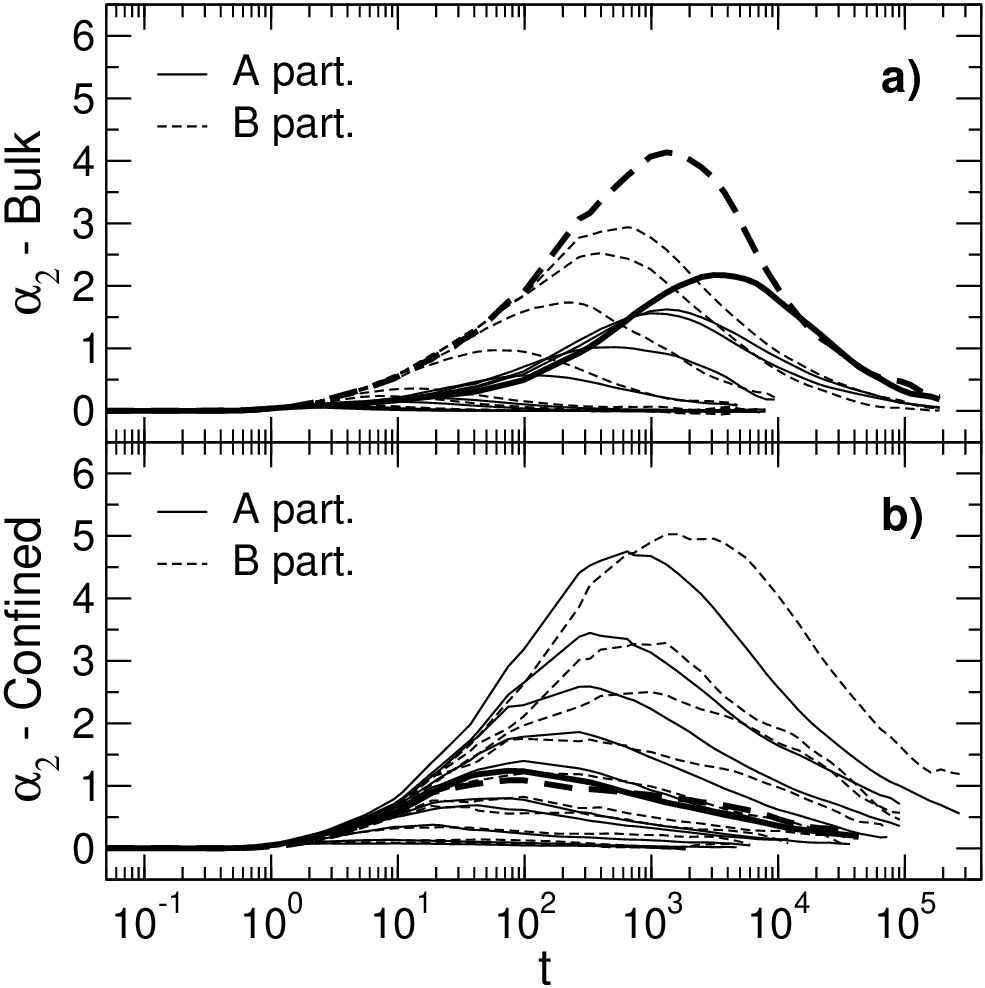}
\end{center}
\caption{The non-Gaussian parameter $\alpha_A^{(2)}(t)$ (continuous lines) 
and $\alpha_B^{(2)}(t)$ (dashed lines) for the bulk (a) and 
confined~\cite{pella1} 
(b) systems. 
Higher curves correspond to lower temperatures $0.48 \le T \le 5.0$ for 
the bulk and $0.37 \le T \le 5.0$ for the confined systems. Thicker lines in 
both figures are associated to the lowest temperature ($T=0.48$) for the bulk.}
\protect\label{fig:13}
\end{figure}

\begin{figure}[h!]
\begin{center}
\includegraphics[width=8.5cm]{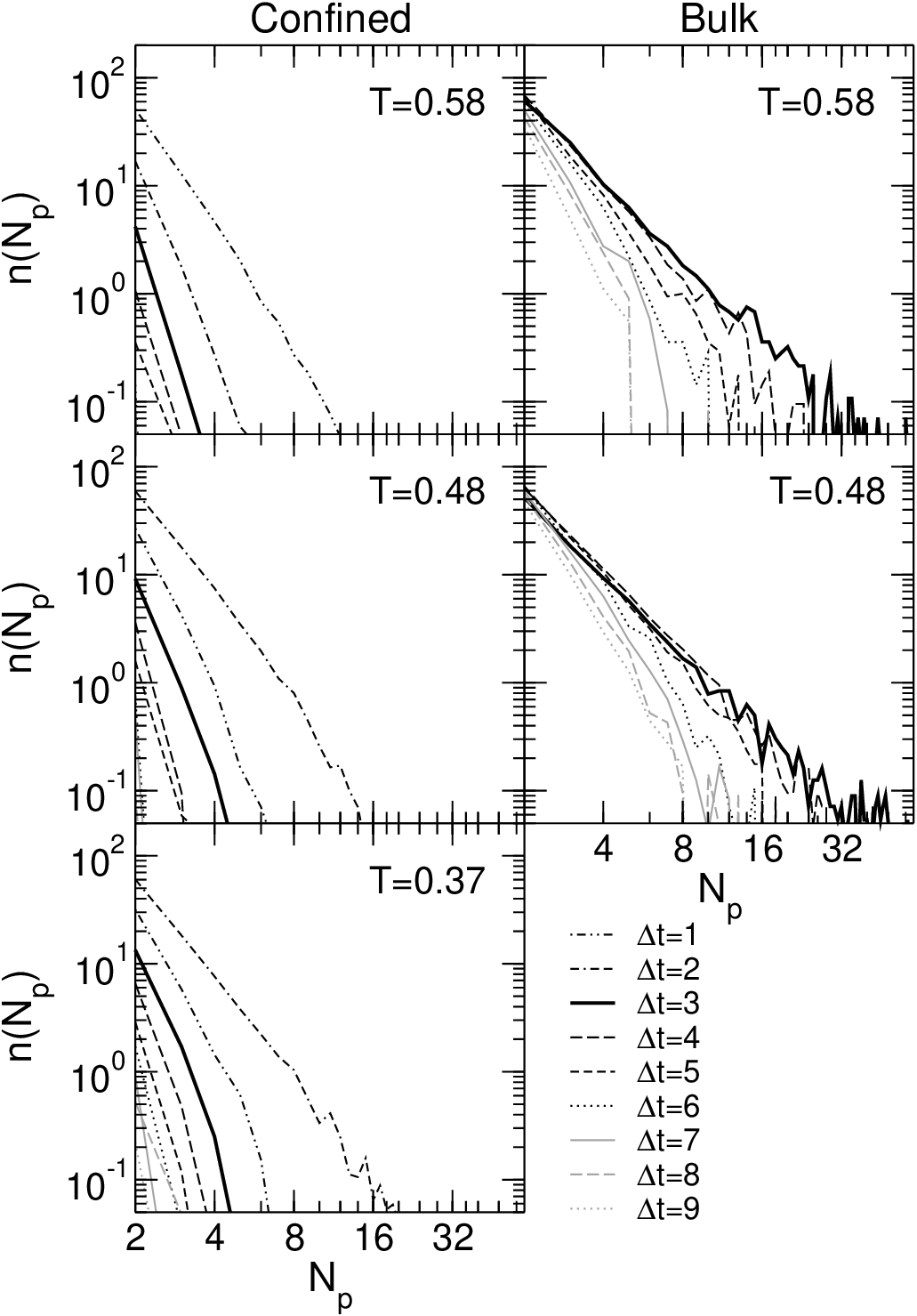}
\end{center}
\caption{Distribution of the size $N_p$ (number of particles) 
of the dynamical clusters found in the confined (first row of panels) 
and in the bulk (second row of panels) liquids, for the temperatures 
$T=0.58$, $0.48$ and $0.37$. Different curves correspond to different 
time ranges used in the sampling (life-time of the clusters).}
\protect\label{fig:14}
\end{figure}

\subsection{Non-Gaussian Parameter}

The deviation from the 
Gaussian behavior of $G^{(s)}_\mu (r,t)$ can be analyzed in terms of 
the {\it non-Gaussian parameter}~\cite{yip}:
\begin{equation}
\alpha^{(2)}=\frac{\left< r^4(t) \right> - 
\left< r^4(t) \right>_{\rm Gauss}}{\left< r^4(t) \right>_{\rm Gauss}} = 
\frac{3\left< r^4(t) \right>}{5\left< r^2
(t) \right>^2}-1
\label{eq:4b}
\end{equation}
which is the relative deviation between the $\left< r^4(t) \right>$
obtained from simulations and
the value it should have if $G^{(s)}(r,t)$ was Gaussian.
The non Gaussian parameter can be studied in order to quantify the
effects of dynamic heterogeneity, as hopping processes, and it 
can also be obtained experimentally from the Q-dependence 
of the Debye Waller factor.~\cite{Zorn}

In figure \ref{fig:13}, $\alpha^{(2)}(t)$ is shown 
as a function of $t$ for A and B particles for different temperatures. 
The same result for the confined
system is reported for comparison~\cite{pella1}. 
We note that the $\alpha^{(2)}(t)$ behavior for both 
systems is similar to that found in several other 
simulations on bulk liquids close to glass transition~\cite{kob}: 
$\alpha^{(2)}(t)$ increases significantly in the late 
$\beta$ relaxation region, near the onset of the $\alpha$ 
relaxation, where it reaches its maximum, then it decays to 
zero as $t\to \infty$. Furthermore, a master curve  
is found overlapping the 
individual curves in a time interval just before they attain 
their maximum. However we note that, in our confined system, 
such master curve is considerably less defined than in the bulk.

In the bulk system the non Gaussian parameter
appears very different for $A$ and $B$ particles and never
exceeds the value of $2$ for $A$ particles. This can be connected
to the observed negligible hopping above $T_C$ for
$A$ particles. In the confined system
we have instead similar parameters for $A$ and $B$ particles
and their values grow substantially as we supercool.
For the lowest temperature investigated in confinement the
peak heights are around $5$. Hopping appears therefore to cause
a much more marked deviation from gaussianity than
MCT behavior and it also seems to lead
to a disappearance of the master curve.

\section{Cluster dynamics}

Extensive investigations of LJBM with a large number of particles
have shown recently the important role of the cooperative molecular motion
in the interpretation of the slow dynamics and the behavior of the
relaxation in liquids upon supercooling close to $T_C$.~\cite{donati}
We expect that confinement induces modifications in the CRR mechanism.
We have investigated for our LJBM both in bulk and in
confined phases the possible existence of a spatially correlated dynamics.
An estimate of the typical distance of the particles correlation
can be obtained from an analysis
of clusters of nearest neighbors.     
In our analysis two particles belong to the same cluster if
the maximum distance they reach from each other within
a fixed time $\Delta t$ is less equal than $\sigma_{\mu \nu} + \Delta \sigma$
where $\Delta \sigma=0.1$. In Fig.~\ref{fig:14} we show the
distribution functions of the size of the number of particles,
$N_p$ of the dynamical clusters for different time ranges $\Delta t$
used in the sampling at different temperatures.

We observe that in the bulk the larger clusters, corresponding to
the shorter fixed times, arrive to circa $32$ particles and
that at the lowest temperature investigated 
the distribution function of the clusters remains unchanged up to $\Delta t=5$.
In confinement the maximum size of the clusters does not exceed
$16$ particles and persists only for a very short time also for the
lowest temperature.
In the bulk a very precise analysis of these spatial correlations
has been already carried out for the same LJBM.
The authors use a much larger sample and are able to distinguish between
mobile and immobile particles.~\cite{donati} 
Both these kind of particles show a distribution of the sizes similar
to what found in our smaller bulk sample.
Comparing our analysis of the bulk and confined systems
we observe that in confinement 
spatial correlations are strongly depressed. This can be
connected to the fact that at variance with the bulk in the 
confined system hopping
mechanisms intervene while cages are still relaxing.
The authors of ref.~\onlinecite{donati} in their study of the bulk phase
of the LJBM underline in fact the
importance of subsequential relaxation processes for spatially
correlated dynamics.
At any given time most particles are localized in cages
and a small fraction forms cooperative strings. Once 
rearranged these particles become caged and others become mobile.
The clarification of this point needs further investigation
and can be possibly connected with the behavior
of the inherent structures in confinement.~\cite{strutin}

\section{Conclusions}

We performed an accurate study of the behavior 
of a LJBM upon supercooling.
The system has been studied along the thermodynamical
path corresponding to the isochoric path of a previous study 
of the same liquid in confinement. In this way we could carry out
a quantitative comparison between the two systems.

The analysis of the density correlators
in the alpha-relaxation region shows
that the MCT predictions are satisfied by our bulk system
that approaches the critical curve upon decreasing temperature
with a path where the volume is not constant.

On comparing the SISF of the bulk with the confined one
we  have confirmed and quantified the differences and analogies 
between the two systems that we did hypothize in our previous 
studies~\cite{euro-noi,pella1,pella2}.
In particular we have
observed a difference in the slope of the tail
of the correlators and a much less defined plateau in confinement.
This leads to a substantially lower value of the stretching parameter $\beta$.
Besides the $\beta$ values do not reach a constant value
upon decreasing temperature at variance with those of the bulk liquid.
The crossover temperature in the confined case is lower with respect to bulk.  

Discrepancies with the MCT predictions for the exponent $\gamma$
determining the asymptotic behavior of the relaxation time and
the diffusion coefficient are found to be more marked in confinement.
This can be ascribed to the hopping phenomenon that in confinement
plays above and near $T_C$ a more significant role than in the bulk.
In the bulk in fact 
hopping effects appear only for the smaller $B$ particles and in 
a time range well distinct from the time range of the cage effect.
In confinement we observe hopping effects for both species of
particles in a time scale that coincides with the time scale of caging. 
As a consequence MCT behavior is partially hidden. In fact
the TTSP is verified also in confinement
but in a much smaller range of temperatures, analogously to the
power law behavior.
Also the prediction that the exponent 
$b$ extracted from the von Schweidler
should be the same for $A$ and $B$ particles
is verified only in confinement.

Hopping effects determine
deviation from gaussianity more marked in confinement than
in bulk. The non Gaussian parameters are different for
$A$ and $B$ particles in the bulk but they are very similar in confinement.   

Finally our comparison of the cluster dynamics in confinement and in bulk
shows that clusters in the confined system have smaller sizes and
last shorter time than in the bulk. Further analysis of cooperativity
in confined glass former would require to study larger systems, but
present results show that in the presence of a restricted 
environment cooperativity is at least partially inhibited.

\end{document}